\documentclass[9pt,twocolumn,twoside]{opticajnl}
\journal{opticajournal} 

\usepackage{graphicx}
\usepackage{dcolumn}
\usepackage{bm}
\usepackage{lineno}
\usepackage{amsthm,amsmath,amssymb}
\usepackage{mathrsfs}
\usepackage{appendix}
\usepackage{amstext} 
\usepackage{url}
\usepackage{soul}
\usepackage{xcolor}
\usepackage{float} 
\usepackage{natbib}
\usepackage{booktabs}
\usepackage{graphicx}
\usepackage{siunitx}
\usepackage{makecell}
\usepackage{multirow} 
\usepackage[normalem]{ulem}
\usepackage{upgreek}
\usepackage{multirow}

\definecolor{Black}{named}{black}

\newcommand{\be}{\begin{equation}}
\newcommand{\ee}{\end{equation}}

\newcommand{\sket}[1]{{\ensuremath{\lvert#1\rangle}}}
\newcommand{\lket}[1]{{\ensuremath{\left\lvert#1\right\rangle}}}
\newcommand{\ket}[1]{\if@display\lket{#1}\else\sket{#1}\fi}

\usepackage{stackengine}
\newcommand{\sbra}[1]{{\ensuremath{\langle#1\rvert}}}
\newcommand{\lbra}[1]{{\ensuremath{\left\langle#1\right\rvert}}}
\newcommand{\bra}[1]{\if@display\lbra{#1}\else\sbra{#1}\fi}

\newcommand{\sketbra}[2]{{\ensuremath{\lvert #1\rangle\!\langle #2\rvert}d}}
\newcommand{\lketbra}[2]{{\ensuremath{\left\lvert #1\right\rangle\!\!\left\langle #2\right\rvert}}}
\newcommand{\ketbra}[2]{\if@display\lketbra{#1}{#2}\else\sketbra{#1}{#2}\fi}

\theoremstyle{plain}

\theoremstyle{definition}

\renewcommand{\copyrightstatement}{}
\doi{}
\newcommand{\zcl}{\textcolor{black}}

\begin{document}





\title{Fiber-integrated Quantum Frequency Conversion for Long-distance Quantum Networking}


\author[1,2,3,4]{Zhichuan Liao}
\author[1,5]{Ao Shen}
\author[1,*]{Lai~Zhou}
\author[4]{Nan Jiang}
\author[1]{Zhiliang~Yuan}

\affil[1]{Beijing Academy of Quantum Information Sciences, Beijing 100193, China}
\affil[2]{Institute of Physics, Chinese Academy of Sciences, Beijing 100190, China}
\affil[3]{University of Chinese Academy of Sciences, Beijing 101408, China}
\affil[4]{School of Physics and Astronomy, Beijing Normal University, Beijing 100875, China}
\affil[5]{National Laboratory of Solid State Microstructures and School of Physics, Collaborative Innovation Center of Advanced Microstructures, Nanjing University, Nanjing 210093, China}
\affil[*]{zhoulai@baqis.ac.cn}

\begin{abstract}


Signal photons emitted by quantum nodes typically fall outside the low-loss telecom window of optical fibers, leading to severe transmission losses.
Quantum frequency conversion (QFC) 
offers an effective optical interface that bridges quantum nodes with telecom-band channels, enabling long-distance quantum communication.
In this work, we demonstrate a compact, fiber-integrated QFC system with low noise and a high single-to-noise ratio (SNR). Using a periodically poled lithium niobate (PPLN) waveguide,  
input photons at 637.2~nm are down-converted to telecom photons at 1588.3~nm. 
Our system achieves a total conversion efficiency of approximately 9\%, with pump-induced noise suppressed to 154~Hz. 
For input photon rates of 32.7, 118.0, and 327.7~kHz, the corresponding SNRs are 12.3, 43.9, and 117.8, respectively. 
We further develop a theoretical model to simulate the entanglement fidelity between nitrogen-vacancy (NV) center spins and the frequency-converted telecom photons.
At the emission rate of an NV center, our QFC system maintains an expected fidelity exceeding 52\% over a transmission distance of 100~km. These findings highlight the potential of our QFC system for scalable, long-distance quantum networking.

\end{abstract}

\maketitle

\section{Introduction}

Quantum networks~\cite{science.aam9288,covey2023quantum,lpor.202100219} comprising distributed quantum nodes have shown great potential for enabling secure communication~\cite{bhaskar2020experimental,zhou2024independent,sun2016quantum,shen2023hertz,fan2025quantum}, distributed quantum computing~\cite{wei2025universal}, and enhanced quantum sensing~\cite{Gottesman2012,malia2022distributed}. Progress in this field has been largely driven by advances in generating entanglement between separated nodes~\cite{Hofmann2012,Bernien2013,Usmani2012,nature.10.1038} and the development of long-lived quantum memories~\cite{bao2012efficient}. 
However, the deployment of a large-scale quantum internet faces practical challenges. 
Signal photons emitted by quantum nodes typically fall outside the low-loss telecom window of optical fibers, leading to significant attenuation over long-distance transmission.
To address this fundamental issue, quantum frequency conversion (QFC)~\cite{ikuta2013high,Zaske2012} has emerged as a pivotal technology, enabling qubit platform operating at a suitable wavelength to interface seamlessly with existing telecom infrastructure. QFC has been applied to convert the wavelength of single photons emitted by quantum dots~\cite{deGreve2012}, atoms~\cite{maring2017photonic,zhang2024fast}, ions~\cite{kucera2024demonstration}, and nitrogen-vacancy \zcl{(NV)} centers in diamond~\cite{PRXQuantum.3.020359}. 
With QFC serving as a critical component, 
recent experiments have demonstrated matter-photon entanglement over distances exceeding 100~km~\cite{zhou2024long1,krutyanskiy2024multimode}.

\begin{figure*}[!t]
\centering
\includegraphics[width=13.5 cm]{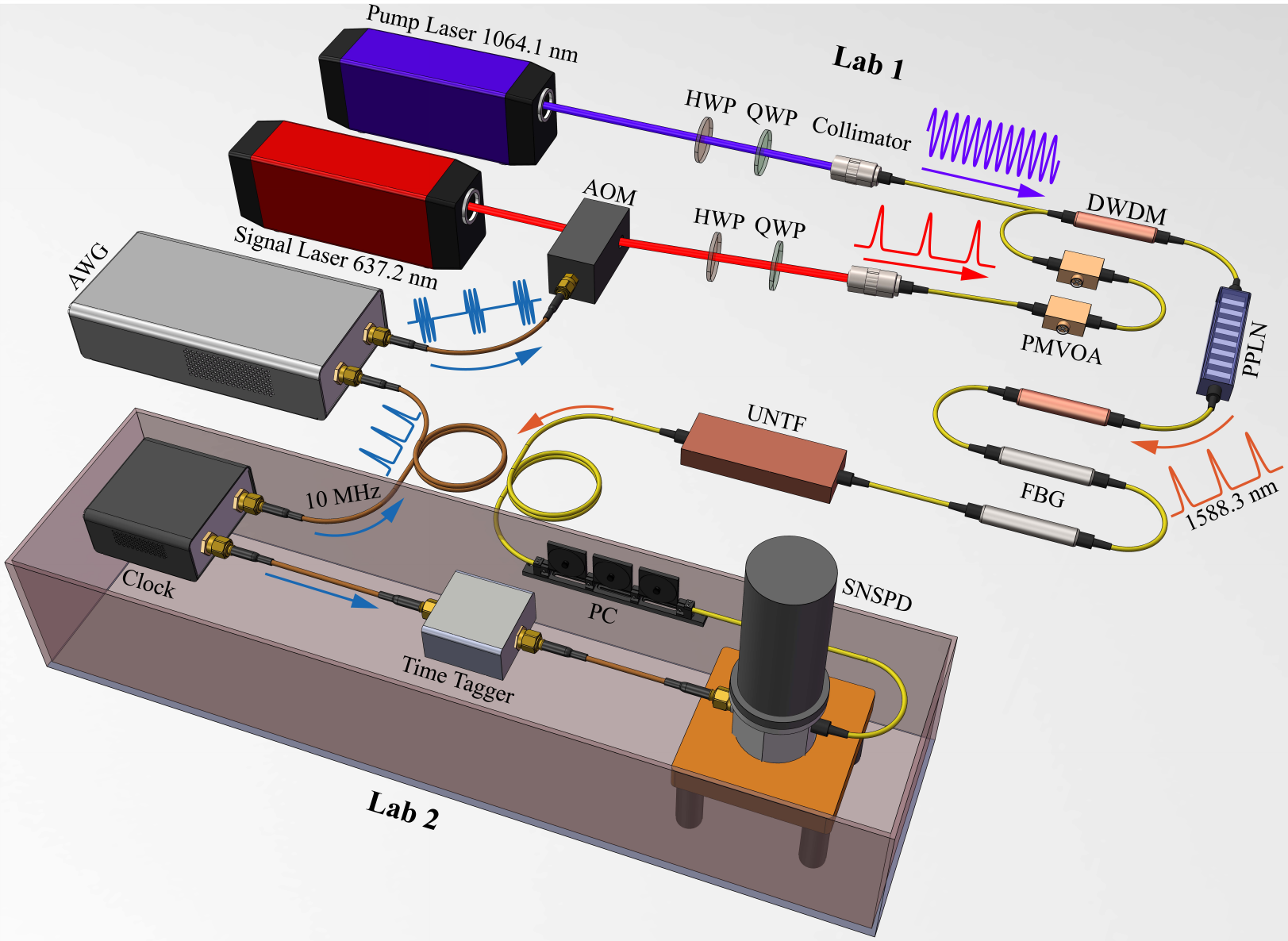}
\caption{Experimental setup. A continuous-wave (CW) laser at 637.2~nm (signal laser) is carved into optical pulses using an acousto-optic modulator (AOM), and then attenuated to single photon level by two cascaded polarization maintaining manual variable optical attenuators (PMVOAs). The attenuated signal pulses are subsequently combined with a 1064.1 nm pump laser  via dense wavelength division multiplexing (DWDM), then coupled into a periodically poled lithium niobate (PPLN) waveguide for frequency conversion. The converted photons are spectrally filtered using a combination of DWDM, fiber Bragg gratings (FBG), and a ultra-narrowband tunable optical filter (UNTF). A clock source, time tagger and  superconducting nanowire single-photon detector (SNSPD) are located in a separate laboratory, where the filtered photons are delivered via optical fiber for single-photon detection. \zcl{HWP, half-wave plate. QWP, quarter-wave plate. AWG, arbitrary waveform generator.} PC, polarization controller.}
\label{fig:setup}
\end{figure*}

QFC can be achieved through sum- and difference-frequency generation in materials exhibiting strong nonlinearities, such as periodically poled lithium niobate (PPLN)~\cite{Strassmann:19,Maring:18} and periodically poled potassium titanyl phosphate (PPKTP)~\cite{Ramelow2012,Rutz2017,PhysRevApplied.20.054010}.
A common challenge in QFC is the noise introduced by strong pump light, arising from processes such as spontaneous parametric down-conversion (SPDC) and Raman scattering~\cite{Pelc:11}. 
The noise photons degrade the signal-to-noise ratio (SNR), thereby limiting the fidelity of matter–photon entanglement and, consequently, constraining the generation of high-quality matter–matter entanglement.
Noise suppression strategies include spectral filtering using narrowband elements such as bandpass filter, etalon, and Fabry–Pérot cavity to reduce broadband noise~\cite{PhysRevLett.120.203601,PhysRevApplied.23.024049,Dreau2018,Murakami:25}; two-stage QFC architectures that employ pump photons at frequencies lower than the target output to eliminate SPDC noise~\cite{qute.202300228,Esfandyarpour:18}; and the use of nonlinear materials with strong phase mismatch to intrinsically suppress unwanted nonlinear processes~\cite{Geus:24}.
Existing QFC systems are often implemented in free-space configurations, which demand precise optical alignment and rely on bulky components, limiting their flexible deployment in field-based quantum networks. It is therefore essential to address both performance and practicality for a QFC system.

In this study, we demonstrate a compact, fiber-integrated QFC system with low noise and high SNR, which converts 637.2~nm signal photons (corresponding to the NV center) into telecom photons at 1588.3~nm using a PPLN waveguide. 
Our system achieves a total conversion efficiency of 9\%, while suppressing pump-induced noise to 154~Hz through a multi-stage filtering module. 
At an input photon count rate of 32.7~kHz, we achieve an SNR of 12.3, marking a 76\% improvement compared to the pioneering setup~\cite{Dreau2018}.
We further develop a theoretical model to evaluate the spin–photon entanglement fidelity over long fiber links.
At the emission rate of an NV center~\cite{Dreau2018}, our QFC system exhibits 
an expected fidelity above 52\% after 100~km of fiber transmission, highlighting its potential for long-distance quantum networking.

\section{Results}\label{sec2}
\subsection{Experimental Setup}\label{subsec1}

As shown in Fig.~\ref{fig:setup}, a 637.2~nm continuous-wave (CW) beam is carved into pulses by an acousto-optic modulator (AOM) and attenuated to the single-photon level. The signal photons are then combined with a 1064.1~nm pump beam and frequency-converted to telecom photons at 1588.3~nm via a PPLN waveguide. 
The 637 nm laser (FL-SF-637-0.1-CW, Precilasers) has a linewidth of 4.4~kHz, \zcl{while} the 1064~nm laser (FL-SF-1064-10-CW, Precilasers)  has a linewidth of 2.3~kHz and supports a maximum output power of 10~W.
The beam polarization is adjusted using a combination of a half-wave plate (HWP) and a quarter-wave plate (QWP) to maximize the frequency-conversion efficiency.
The converted photons are subsequently passed through a multi-stage filtering module to suppress noise and transmitted via optical fiber to a separate laboratory, where the photons are detected using a superconducting nanowire single-photon detector. The entire frequency-conversion module is fully fiber-coupled, ensuring high stability and alignment-free operation.

Driven by an arbitrary waveform
generator (AWG), the AOM carves the 637~nm beam into 300~ns long pulses at a repetition rate of 1~MHz, corresponding to a duty cycle of 30\%. The optical power of the 637~nm pulses is measured to be approximately 1.74~mW. To reach the single-photon level matching the typical zero-phonon-line (ZPL) emission rate of \zcl{a} NV center, with a photon count rate on the order of tens of kilohertz~\cite{science.1253512}, the 637.2~nm pulses are attenuated using two cascaded polarization-maintaining variable optical attenuators (PMVOAs). By introducing a total attenuation of up to  117.6~dB, the 637.2~nm pulses are reduced to the desired single-photon level.


The PPLN waveguide adopts a lithium-niobate-on-insulator (LNOI) ridge structure, which provides a high refractive-index contrast between the core and the cladding~\cite{Lu:22}. The enhancement of the nonlinear interaction enables highly efficient frequency conversion within a centimeter-scale waveguide and at relatively low pump power. The input and output ports of the PPLN waveguide are coupled to PM980 and PM1550 fibers, respectively. Considering both the fiber coupling losses and the propagation loss within the waveguide, the overall transmission efficiency is approximately 36\% for the converted photons.
The waveguide is 20~mm in length, \zcl{with a poling period of $11.85~\upmu \rm{m}$}, and is temperature-stabilized at $33.80\ ^{\circ}\rm{C}$ using a thermoelectric controller with a stability of $\pm 0.01 \ ^{\circ}\rm{C}$, ensuring optimal phase matching.

Given that the pump wavelength is shorter than the target telecom wavelength and the spectral separation is substantial, the dominant noise is attributed to SPDC photons~\cite{Pelc2010}. To mitigate the effect of noise, we introduce a hybrid filtering module that integrates a \zcl{dense wavelength division multiplexing (DWDM)}, two fiber Bragg gratings (FBGs), and an ultra-narrowband tunable filter (UNTF) to achieve significant noise suppression.
The DWDM effectively removes the residual pump light, while the two 10 GHz-bandwidth FBGs provide an initial stage in suppressing the broadband SPDC noise, and the additional 250 MHz-bandwidth UNTF is employed to further enhance the noise suppression.
The overall filtering module provides over 97~dB of noise suppression, while introducing approximately 4.4~dB of loss to the converted telecom photons.

\begin{figure}[t]
\centering
\includegraphics[width=1.02\linewidth]{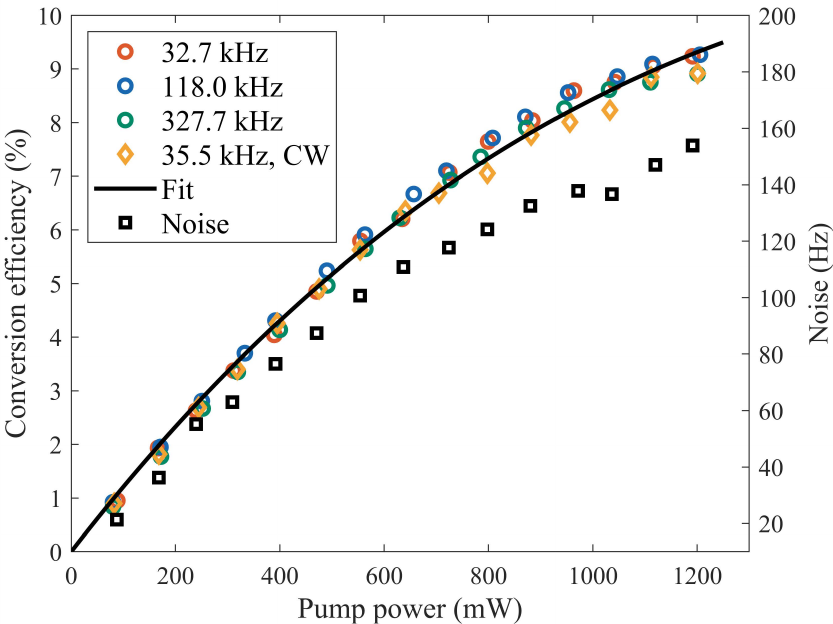}
\caption{Conversion efficiency and noise versus pump power. The conversion efficiency for pulsed (circles) and CW (yellow diamonds) signal light is plotted as a function of pump power (left axis). The black solid line denotes a fit to \eqref{eq1}. The corresponding noise count rate is shown on the right axis (black squares).
}
\label{fig:2}
\end{figure}

After filtering, the telecom photons are routed via fiber to a separate laboratory, where single-photon detection is performed using a superconducting nanowire single-photon detector (SNSPD). The fiber transmission loss between the two laboratories is approximately 1.2~dB. 
By optimizing the polarization state with a polarization controller (PC), the SNSPD operates at a detection efficiency of 90\% and a dark count rate of 54~Hz. A clock source distributes a 10~MHz signal to the AWG as a reference to remotely synchronize the pulse generation and detection modules.
The time tagger operates with a 10~ns time bin and a 300~ns gating window to temporally separate the telecom photons from noise.


\subsection{Conversion efficiency and SNR}\label{subsec2}

\begin{figure}[t]
\centering
\includegraphics[width=0.97\linewidth]{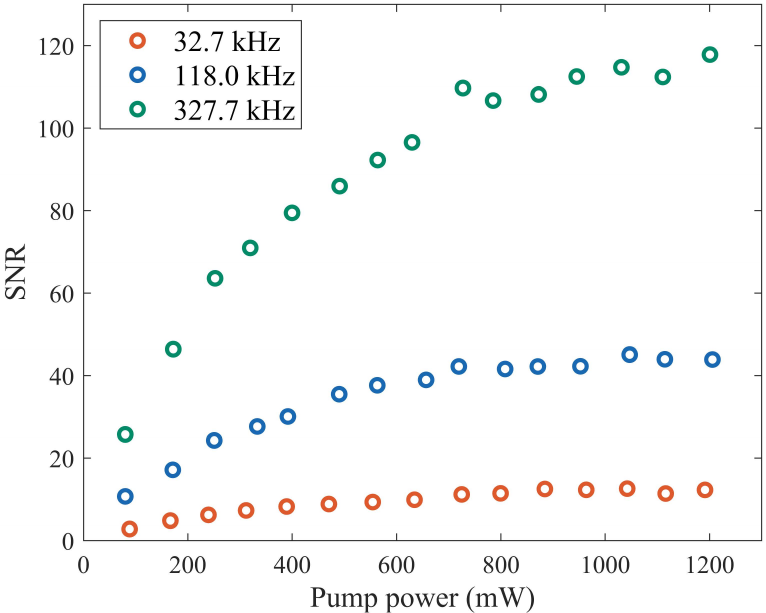}
\caption{Signal-to-noise ratio of the converted signal as a function of pump power. The circles represent the measured SNRs for input photon count rates of 32.7~kHz (red), 118.0~kHz (blue), and 327.7~kHz (green) at 637~nm.}
\label{fig:3}
\end{figure}

The conversion efficiency and corresponding noise are characterized as functions of pump power, as shown in Fig.~\ref{fig:2}.
The measured conversion efficiency includes the loss arising from the combined DWDM, fiber-waveguide coupling, and hybrid filtering module.
The CW signal light, with a photon count rate of 35.5~kHz, yields a maximum conversion efficiency of 8.92\% at a pump power of 1.2~W.
The pulsed signals, attenuated by 117.6, 112.0, and 107.5~dB to photon count rates of 32.7, 118.0, and 327.7~kHz, respectively, achieve maximum conversion efficiencies of 9.24\%, 9.27\%, and 8.92\% at the same pump power, as indicated by the red, blue, and green circles in Fig.~\ref{fig:2}.
The measured conversion efficiencies are fitted using the following formula~\cite{Pelc2010}
\begin{equation}
\label{eq1}
\eta_c(P)=\eta_{max}{\rm sin}^{2}\left(L\sqrt{\alpha_{QFC}P}\right),
\end{equation}
where $L=20\ {\rm mm}$ is the length of the waveguide, $P$ is the pump power, $\alpha_{QFC}=2.87\times10^{3}\ {\rm W^{-1}m^{-2}}$ is the normalized power efficiency determined by the PPLN waveguide, $\eta_{max}=10.95\%$ is the 
simulated maximum conversion efficiency taking into account all propagation losses.
The fitted black curve shows excellent agreement with the experimental results. It is worth noting that the measured maximum  $\eta_c \approx 9\%$ corresponds to a high internal conversion efficiency of approximately 80\% within the waveguide. The primary reduction in $\eta_c$ originates from the propagation efficiencies of the combined DWDM (87\%), PPLN waveguide (36\%), and filtering module (36\%).
While free-space configurations typically exhibit higher propagation efficiencies in the PPLN waveguide (above 70\%~\cite{Strassmann:19,Dreau2018,PhysRevApplied.23.024049}), our fiber-integrated system achieves a balanced trade-off between performance and flexibility.

\begin{table}[t]
\caption{Expected \zcl{SNRs and} fidelities for different fiber lengths when $R_{S}=32.7/35.5\ \rm{kHz}$, extracted from the results shown in Fig.~\ref{fig:4}.}\label{table1}
\begin{tabular*}{\linewidth}{@{\extracolsep\fill}lcccccc}
\toprule%
& \multicolumn{3}{c}{Our work} & \multicolumn{3}{c}{Ref.~\cite{Dreau2018}} \\
\cmidrule{2-4}\cmidrule{5-7}
$R_{S}$ (kHz) & \multicolumn{3}{c}{32.7} & \multicolumn{3}{c}{35.5}\\
\midrule
\zcl{Conversion} & \multicolumn{3}{c}{\multirow{2}{*}{\zcl{9}}} & \multicolumn{3}{c}{\multirow{2}{*}{\zcl{17}}}\\
\zcl{efficiency (\%)}\\
\midrule
Length (km) & 0 & 60 & 100 & 0 & 60 & 100 \\
\midrule
\zcl{SNR} & \zcl{13.74} & \zcl{4.19} & \zcl{1.16} & \zcl{6.87} & \zcl{1.30} & \zcl{0.32} \\
\midrule
Fidelity (\%) & 90.5 & 75.8 & 52.5 & 83.1 & 54.5 & 35.3\\
\midrule
\end{tabular*}
\end{table}

The noise count rate, shown by the black squares, increases linearly with pump power and remains as low as 154~Hz at 1.2~W.
It is worth noting that the reported noise does not include the dark count rate in the SNSPD. 
Such a low noise level highlights the effectiveness of the hybrid filtering module in suppressing broadband noise contributions. The influence of noise on the signal can be further mitigated through temporal filtering.
Enabled by time-synchronized photon detection, we evaluate the SNR within a 300~ns temporal window (see Methods and Supplementary Note 2 for details).
As shown in Fig.~\ref{fig:3}, the SNR initially increases with pump power due to the enhancement in conversion efficiency. At a pump power of 1.2~W, the SNRs are measured to be 12.3, 43.9, and 117.8, corresponding to input photon count rates of 32.7, 118.0, and 327.7~kHz at 637~nm, respectively. We find that, at a count rate of 32.7~kHz, further increases in pump power can lead to a gradual reduction in SNR, as the additional noise contributions begin to outweigh the gain in conversion efficiency.
Nevertheless, our system demonstrates a noticeable improvement of 76\% in SNR at the typical emission rate of an NV center~\cite{Dreau2018}.
Furthermore, higher input count rates lead to a significant enhancement in SNR, indicating that the ZPL photon count rate from the NV center is a key factor influencing system performance.

\subsection{Entanglement fidelity }\label{subsec3}

\begin{figure}[t]
\centering
\includegraphics[width=1\linewidth]{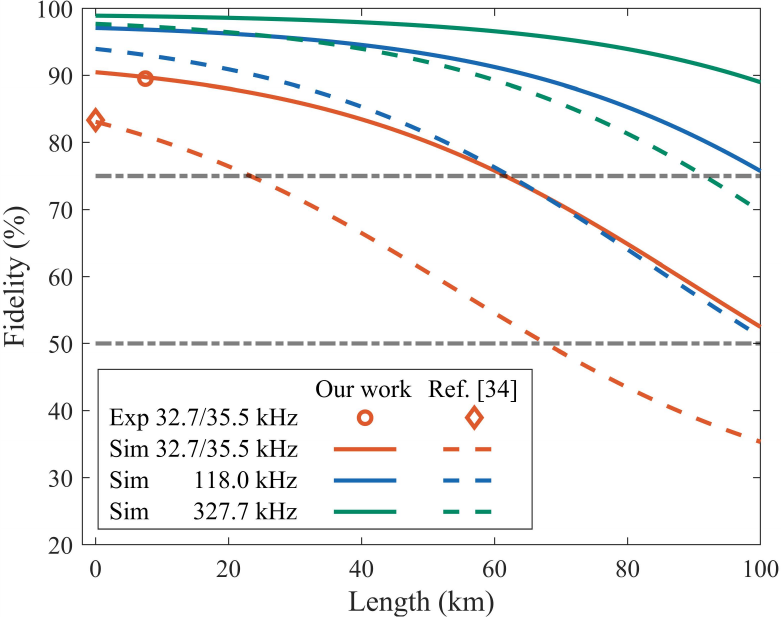}
\caption{Simulated entanglement fidelity as a function of fiber length. The curves are obtained from a theoretical model under two parameter sets: our experimental results (solid lines) and those from Ref.~\cite{Dreau2018} (dashed lines). The circle and diamond are calculated using the measured SNRs. The red solid line and circle correspond to a photon count rate of $R_{S}=32.7\ {\rm kHz}$ (our work), and the red dashed line and diamond correspond to $R_{S}=35.5\ {\rm kHz}$ (Ref.~\cite{Dreau2018}). The blue and green curves represent $R_{S}=118.0\ {\rm kHz}$ and $R_{S}=327.7\ {\rm kHz}$, respectively. The circle marks an equivalent fiber length of 7.5~km, corresponding to 1.2~dB transmission loss between the two laboratories.}
\label{fig:4}
\end{figure}

We \zcl{develop} a theoretical model to evaluate the spin-photon entanglement fidelity as a function of the SNR.
We assume that the fidelity is obtained by performing correlated measurements between the NV center’s electronic spin and the photon in the $X$, $Y$, and $Z$ bases~\cite{PhysRevLett.123.063601}.
The fidelity of the Bell state $\left|\psi_{+}\right>$ is given by $F=\left(1+ V_{X}+ V_{Y}-V_{Z}\right)/4$, where $V_{a}\ \left(a=X,Y,Z\right)$ denotes the visibility measured in the corresponding basis~\cite{doi:10.1126/sciadv.adp6442}.
The fidelity after frequency conversion and long-distance transmission can be theoretically derived as (see Supplementary Note 1),
\begin{equation}
\label{eq2}
F=1-\frac{3\left(\eta_{l}\eta_{d}R_{N}+R_{D}\right)}{2\eta_{c}\eta_{l}\eta_{d}R_{S}+4\left(\eta_{l}\eta_{d}R_{N}+R_{D}\right)},
\end{equation}
where $R_{S}$ is the input photon count rate at 637~nm, 
$R_{N}$ is the noise count rate introduced from frequency conversion, $R_{D}$ denotes the dark count rate in the detector, $\eta_{c}$ is the conversion efficiency, $\eta_{d}$ is the detection efficiency, and $\eta_{l}$ denotes the transmission efficiency through the long fiber.
The SNR is expressed as, 
\begin{equation}
\begin{split}
\label{eq3}
{\rm SNR} =\frac{R_{S} \cdot \eta_{c} \eta_{l}\eta_{d}}{R_{N} \cdot \eta_{l}\eta_{d}+R_{D}}.
\end{split}
\end{equation}
Accordingly, the fidelity can be derived as a function of SNR,
\begin{equation}
\label{eq4}
F=1-\frac{3}{2\left({\rm SNR}+2\right)}.
\end{equation}

As shown in Fig.~\ref{fig:4}, the post-conversion fidelity is simulated as a function of fiber length at different  photon count rates $R_{S}$.
Two sets of parameters are considered. The first set corresponds to our experimental results, for example, $R_{N}=154\ {\rm Hz}$ and $\eta_{c}\approx 9\%$. 
The second set is based on the experimental parameters reported in Ref.~\cite{Dreau2018}, with $R_{N}=415\ {\rm Hz}$, $R_{D}=190\ {\rm Hz}$, $\eta_{c}\approx 17\%$, and $\eta_{d}\approx 41\%$.
The transmission efficiency over a fiber length of $l\ \rm{km}$ can be expressed as $\eta_{l}=10^{-0.16l/10}$, assuming an attenuation coefficient of 0.16~dB/km. 
The expected fidelities in our experiment are calculated to be 89.5\% (red circle), 96.7\%, and 98.7\%, based on the measured SNR at $R_{S}$ of 32.7, 118.0, and 327.7~kHz, respectively.
For comparison, a fidelity of 83.3\% (red diamond) can be achieved when ${\rm SNR}\approx7$ and $R_{S}=35.5\ {\rm kHz}$~\cite{Dreau2018}.
Our expected fidelity improves by 6.2\% thanks to the enhanced SNR.

The advantage of a higher SNR becomes more pronounced for long-distance transmission. As shown in Table.~\ref{table1}, 
at a distance of 60~km, our expected fidelity exceeds that of Ref.~\cite{Dreau2018} by about 21.3\%. At 100~km, our expected fidelity remains as high as 52.5\%, whereas that based on Ref.~\cite{Dreau2018} decreases to 35.3\%.
Moreover, for $R_{s}=118.0\ \rm{kHz}$ (blue lines), our expected fidelity remains as high as 75.7\% after 100~km transmission, compared with 51.0\% for the  parameters in Ref.~\cite{Dreau2018}. These results clearly demonstrate the critical role of SNR in maintaining entanglement fidelity over long-distance fiber transmission.

\section{Discussion}\label{sec3}

We demonstrate a compact, fiber-integrated QFC system featuring low noise and high SNR with a PPLN waveguide, and a multi-stage narrowband filtering module effectively suppresses both residual pump light and SPDC noise. 
For input photon count rates of 32.7, 118.0, and 327.7~kHz, the system achieves total conversion efficiencies of 9.24\%, 9.27\%, and 8.92\%, respectively.
At 32.7~kHz, the measured SNR reaches 12.3, representing an improvement of approximately 76\% compared with Ref.~\cite{Dreau2018}. 
We further develop a theoretical model to simulate the entanglement fidelity between NV center spins and frequency-converted photons. At the emission rate of an NV center, our QFC
system maintains an expected fidelity of 52.5\% after 100~km of fiber transmission. 
These findings highlight the potential of our QFC system for scalable long-distance quantum networking.

Additionally, the PPLN waveguide and filtering components in our setup are fiber-coupled, making the system compact, robust, and compatibility with existing fiber infrastructures. 
In contrast, most QFC systems reported so far rely on free-space configurations.
For instance, cavity-based filtering typically requires bulky optical resonators and additional stabilization components~\cite{PhysRevApplied.23.024049}. Meanwhile, although monocrystalline KTA crystals~\cite{Geus:24} can strongly suppress SPDC noise due to large phase mismatch, their relatively low effective nonlinearity necessitates pump powers of several hundred watts, rendering such schemes less suitable for field deployment.

Future directions should focus on improving the waveguide coupling efficiency and increasing the input photon count rate. The current fiber-to-waveguide coupling efficiency of 36\% is considerably lower than the $\textgreater70\%$ typically obtained in free-space implementations. 
The coupling efficiency can be further improved using advanced mode-matching optics or integrated spot-size converters~\cite{Hu:21}.
Furthermore, the input photon count rate $R_{s}$ plays a crucial role in maintaining fidelity over long distances. As shown in Fig.~\ref{fig:4}, at 100~km, the expected fidelities are 52.5\%, 75.7\%, and 89.0\% for $R_{s}=32.7$, 118.0, and 327.7~kHz, respectively. 
However, ZPL photons constitute only about 3–4\% of the total emission from NV centers~\cite{PhysRevX.7.031040}. Strategies such as enhancing the ZPL branching ratio via the Purcell effect, or improving photon collection efficiency through optimized optical designs~\cite{qute.202200142}, could increase the ZPL photon count rate and thereby improve the achievable entanglement fidelity over long-distance links.

\section{Methods}

\subsection{Temporal filtering scheme}

A temporal filtering scheme is employed to discriminate signal photons from background noise. The AOM is driven by an RF signal with a carrier frequency of 82~MHz and a repetition rate of 1~MHz. With a duty cycle of 30\%, this produces optical pulses with a width of 300~ns at 1~$\mu s$ intervals. The SNR is calculated as $(k_{S}-k_{N})/k_{N}$, where $k_{S}$ and $k_{N}$ denote the accumulated photon counts in the signal and noise regions, respectively, within the 300~ns temporal gating window.

\section{Data availability}
All data needed to evaluate the conclusions of this work are presented in the manuscript.
Additional data related to this work are available from the corresponding author on reasonable request.


\bibliography{ref.bib}

\vspace{0.5cm}
\noindent Fig. 1. Experimental setup. A continuous-wave (CW) laser at 637.2~nm (signal laser) is carved into optical pulses using an acousto-optic modulator (AOM), and then attenuated to single photon level by two cascaded polarization maintaining manual variable optical attenuators (PMVOAs). The attenuated signal pulses are subsequently combined with a 1064.1 nm pump laser  via dense wavelength division multiplexing (DWDM), then coupled into a periodically poled lithium niobate (PPLN) waveguide for frequency conversion. The converted photons are spectrally filtered using a combination of DWDM, fiber Bragg gratings (FBG), and a ultra-narrowband tunable optical filter (UNTF). A clock source, time tagger and  superconducting nanowire single-photon detector (SNSPD) are located in a separate laboratory, where the filtered photons are delivered via optical fiber for single-photon detection. HWP, half-wave plate. QWP, quarter-wave plate. AWG, arbitrary waveform generator. PC, polarization controller.

\vspace{0.5cm}
\noindent Fig. 2. Conversion efficiency and noise versus pump power. The conversion efficiency for pulsed (circles) and CW (yellow diamonds) signal light is plotted as a function of pump power (left axis). The black solid line denotes a fit to \eqref{eq1}. The corresponding noise count rate is shown on the right axis (black squares).

\vspace{0.5cm}
\noindent Fig. 3. Signal-to-noise ratio of the converted signal as a function of pump power. The circles represent the measured SNRs for input photon count rates of 32.7~kHz (red), 118.0~kHz (blue), and 327.7~kHz (green) at 637~nm.

\vspace{0.5cm}
\noindent Fig. 4. Simulated entanglement fidelity as a function of fiber length. The curves are obtained from a theoretical model under two parameter sets: our experimental results (solid lines) and those from Ref.~\cite{Dreau2018} (dashed lines). The circle and diamond are calculated using the measured SNRs. The red solid line and circle correspond to a photon count rate of $R_{S}=32.7\ {\rm kHz}$ (our work), and the red dashed line and diamond correspond to $R_{S}=35.5\ {\rm kHz}$ (Ref.~\cite{Dreau2018}). The blue and green curves represent $R_{S}=118.0\ {\rm kHz}$ and $R_{S}=327.7\ {\rm kHz}$, respectively. The circle marks an equivalent fiber length of 7.5~km, corresponding to 1.2~dB transmission loss between the two laboratories.

\section{Acknowledgements}
This work was supported by the National Natural Science Foundation of China (No. 92476116), Beijing Municipal Natural Science Foundation (No. Z230005), and Quantum Science and Technology-National Science and Technology Major Project (No. 2024ZD0302500).

\section{Author Contributions}
L.Z. conceived the project. Z.C.L., L.Z., A.S. performed the experiments. Z.C.L.  processed the data and derived the theoretical model. L.Z., Z.C.L., N.J., Z.Y. analyzed the results. Z.C.L. and L.Z. prepared the manuscript. All authors discussed, improved and approved the manuscript. Z.Y. supervised the project.

\section{Competing interests}
The authors declare no competing interests.

\vspace{1cm}

\textbf{Correspondence} and requests for materials should be addressed to Lai Zhou.


\end{document}


\title{Supplementary Information: Fiber-integrated Quantum Frequency Conversion for Long-distance Quantum Networking}


\author {Zhichuan Liao}
\affiliation{Beijing Academy of Quantum Information Sciences, Beijing 100193, China} 
\affiliation{Institute of Physics, Chinese Academy of Sciences, Beijing 100190, China}
\affiliation{University of Chinese Academy of Sciences, Beijing 101408, China}
\affiliation{School of Physics and Astronomy, Beijing Normal University, Beijing 100875, China}

\author{Ao Shen}
\affiliation{Beijing Academy of Quantum Information Sciences, Beijing 100193, China} 
\affiliation{National Laboratory of Solid State Microstructures and School of Physics, Collaborative Innovation Center of Advanced Microstructures, Nanjing University, Nanjing 210093, China}

\author{Lai Zhou}\email{zhoulai@baqis.ac.cn}
\affiliation{Beijing Academy of Quantum Information Sciences, Beijing 100193, China} 

\author{Nan~Jiang}
\affiliation{School of Physics and Astronomy, Beijing Normal University, Beijing 100875, China}

\author{Zhiliang Yuan}
\affiliation{Beijing Academy of Quantum Information Sciences, Beijing 100193, China}


\maketitle



\subsection*{Supplementary Note 1: Theoretical model for the fidelity of spin-Photon entangled states after Frequency Conversion}


Three energy levels of the nitrogen-vacancy (NV) center electron spin are typically involved: $\left|\uparrow\right>_{g}$, $\left|\downarrow\right>_{g}$, and $\left|\uparrow\right>_{e}$. The electron spin qubit is encoded in the ground states $\left|\uparrow\right>_{g}$ and $\left|\downarrow\right>_{g}$. Assuming the electron spin is initially prepared in the $\left|\uparrow\right>_{g}$ state, it can be excited to the $\left|\uparrow\right>_{e}$ state using a laser, after which it spontaneously emits a zero-phonon-line (ZPL) photon.
The NV-photon entanglement is typically generated using time-bin encoding \cite{tchebotareva2019entanglement}. The electron spin is first prepared in the superposition state $(\left|\uparrow\right>_{g}+\left|\downarrow\right>_{g})/\sqrt{2}$. An optical excitation then promotes the electron to the excited state, leading to the emission of an early time-bin photon $\left|E\right>$. A microwave pulse is subsequently applied to coherently interchange the $\left|\uparrow\right>_{g}$ and $\left|\downarrow\right>_{g}$ spin states. A second optical excitation follows, generating a later time-bin photon $\left|L\right>$. 
This sequence results in the spin–photon entangled state
\begin{equation}
\label{eq.s1-1}
(\left|\downarrow\right>_{g}\left|E\right>+\left|\uparrow\right>_{g}\left|L\right>)/\sqrt{2},
\end{equation}
where the electron spin state is correlated with the photon's temporal mode. This time-bin encoding is robust against fiber dispersion and constitutes a key resource for long-distance quantum networking with NV centers.

When characterizing the fidelity, the NV center spin state must be read out conditioned on the detection of a signal photon by a single-photon detector, enabling coincidence measurements. For clarity, we take $\left|\uparrow\right>_{g}$ and $\left|\downarrow\right>_{g}$ as the eigenstates of the Z basis and denote them as $\left|Z\right>\equiv\left|\uparrow\right>_{g}$ and $\left|-Z\right>\equiv\left|\downarrow\right>_{g}$. The corresponding eigenstates of the X and Y bases are then defined as follows:
\begin{equation}
\left\{
\begin{aligned}
&\left|X\right>\equiv(\left|\uparrow\right>_{g}+\left|\downarrow\right>_{g})/\sqrt{2}, \\
&\left|-X\right>\equiv(\left|\uparrow\right>_{g}-\left|\downarrow\right>_{g})/\sqrt{2}, \\
&\left|Y\right>\equiv(\left|\uparrow\right>_{g}+i\left|\downarrow\right>_{g})/\sqrt{2}, \\
&\left|-Y\right>\equiv(\left|\uparrow\right>_{g}-i\left|\downarrow\right>_{g})/\sqrt{2}.
\end{aligned}\right.\label{eq:Delay_BSM_4user}
\end{equation}



The eigenstates of the X, Y, and Z bases for the photon are defined as follows,
\begin{equation}
\left\{
\begin{aligned}
&\left|Z\right>\equiv\left|E\right>, \\
&\left|-Z\right>\equiv\left|L\right>, \\
&\left|X\right>\equiv(\left|E\right>+\left|L\right>)/\sqrt{2}, \\
&\left|-X\right>\equiv(\left|E\right>-\left|L\right>)/\sqrt{2}, \\
&\left|Y\right>\equiv(\left|E\right>+i\left|L\right>)/\sqrt{2}, \\
&\left|-Y\right>\equiv(\left|E\right>-i\left|L\right>/\sqrt{2}.
\end{aligned}\right.\label{eq:Delay_BSM_4user}
\end{equation}
During the coincidence measurement, we denote $n_{a,b}$ as the number of coincidence events in which the NV center’s electron spin is measured in state  $\left|a\right>$ and the photon is measured in state $\left|b\right>$.




When the NV center is initialized in the $\left|Z\right>$ state, we assume that the number of collected ZPL photons (i.e., signal photons) within a time window $\tau$ is $N_{S}$.
For a coincidence measurement in the Z basis, the entangled spin–photon state $(\left|Z,-Z\right>+\left|-Z,Z\right>)/\sqrt{2}$ yields the following coincidence counts: $n_{Z,Z}= n_{-Z,-Z} = 0,~n_{Z,-Z}= n_{-Z,Z} = N_{S}/2$. After frequency conversion, spontaneous parametric down-conversion (SPDC) noise photons are generated and mix with the converted signal photons.
Let $N_{N}$ denote the number of SPDC noise photons. 
The corresponding coincidence counts are then $n_{Z,Z}=n_{-Z,-Z}=N_{N}/2, n_{Z,-Z}=n_{-Z,Z}=(\eta_{c}N_{S}+N_{N})/2$, where $\eta_{c}$ denotes the frequency conversion efficiency. Similarly, the contribution of SPDC noise to the coincidence counts in the X and Y bases is summarized in Supplementary Table~\ref{table1}.

\begin{table}[h]
\caption{Coincidence counts in different measurement bases after frequency conversion.}
\label{table1}
\begin{tabular*}{0.7\linewidth}{@{\extracolsep\fill}llllccccc}
\toprule%
& $n_{a,a}$ & $n_{a,-a}$ & $n_{-a,a}$ & $n_{-a,-a}$\\
\midrule
$a=X$ & $\dfrac{\eta_{c}N_{S}}{2}+N_{N}$ & $0$ & $N_{N}$ & $\dfrac{\eta_{c}N_{S}}{2}$\\
\midrule
$a=Y$ & $\dfrac{\eta_{c}N_{S}}{2}+\dfrac{N_{N}}{2}$ & $\dfrac{N_{N}}{2}$ & $\dfrac{N_{N}}{2}$ & $\dfrac{\eta_{c}N_{S}}{2}+\dfrac{N_{N}}{2}$\\
\midrule
$a=Z$ & $\dfrac{N_{N}}{2}$ & $\dfrac{\eta_{c}N_{S}}{2}+\dfrac{N_{N}}{2}$ & $\dfrac{\eta_{c}N_{S}}{2}+\dfrac{N_{N}}{2}$ & $\dfrac{N_{N}}{2}$\\
\midrule
\end{tabular*}
\end{table}

Next, we consider fiber transmission loss and single-photon detection. The fiber transmission factor is given by $\eta_{l}=10^{-\alpha L/10}$, where $L$ is the fiber length in kilometers and $\alpha$ is the attenuation coefficient in dB/km. $\eta_{d}$ denotes the detection efficiency of the single-photon detector, and $N_{D}$ is the number of dark counts within the time window $\tau$. 
Under these conditions, the number of detected signal photons becomes $\eta_{c}\eta_{l}\eta_{d}N_{S}$, while the number of detected SPDC noise photons becomes $\eta_{l}\eta_{d}N_{N}$. 
Note that the coincidence measurement in each basis acquires an additional count of $N_{D}/2$ due to dark counts.

The entanglement fidelity of the Bell state is given by~\cite{Wagenknecht2010,doi:10.1126/sciadv.adp6442},
\begin{equation}
F\left(\left|\psi_{\pm}\right>\right)=\left(1\pm V_{X}\pm V_{Y}-V_{Z}\right)/4,
\end{equation}
where $\left|\psi_{\pm}\right>=\left(\left|01\right>\pm\left|10\right>\right)/\sqrt{2}$, and $V_a$ ($a=X,Y,Z$) is the visibility in each basis, 
\begin{equation}
\label{eq.s3}
V_{a}=\frac{n_{a,a}-n_{a,-a}-n_{-a,a}+n_{-a,-a}}{n_{a,a}+n_{a,-a}+n_{-a,a}+n_{-a,-a}}, (a=X,Y,Z).
\end{equation}
Hence, the theoretical model for the fidelity is derived as
\begin{equation}
\label{eq.s7}
F=1-\frac{3(\eta_{l}\eta_{d}N_{N}+N_{D})}{2\eta_{c}\eta_{l}\eta_{d}N_{S}+4(\eta_{l}\eta_{d}N_{N}+N_{D})}.
\end{equation}
Similarly, the fidelity can be expressed as a function of the photon count rate as follows,
\begin{equation}
\label{eq.s7}
F=1-\frac{3(\eta_{l}\eta_{d}R_{N}+R_{D})}{2\eta_{c}\eta_{l}\eta_{d}R_{S}+4(\eta_{l}\eta_{d}R_{N}+R_{D})},
\end{equation}
where $R_{S}$ denotes the ZPL photon count rate, $R_{N}$ the SPDC noise count rate, and $R_{D}$ the dark counts rate.

\subsection*{Supplementary Note 2: Temporal filtering scheme}

\begin{figure}[htbp]
\centering
\includegraphics[width=\linewidth]{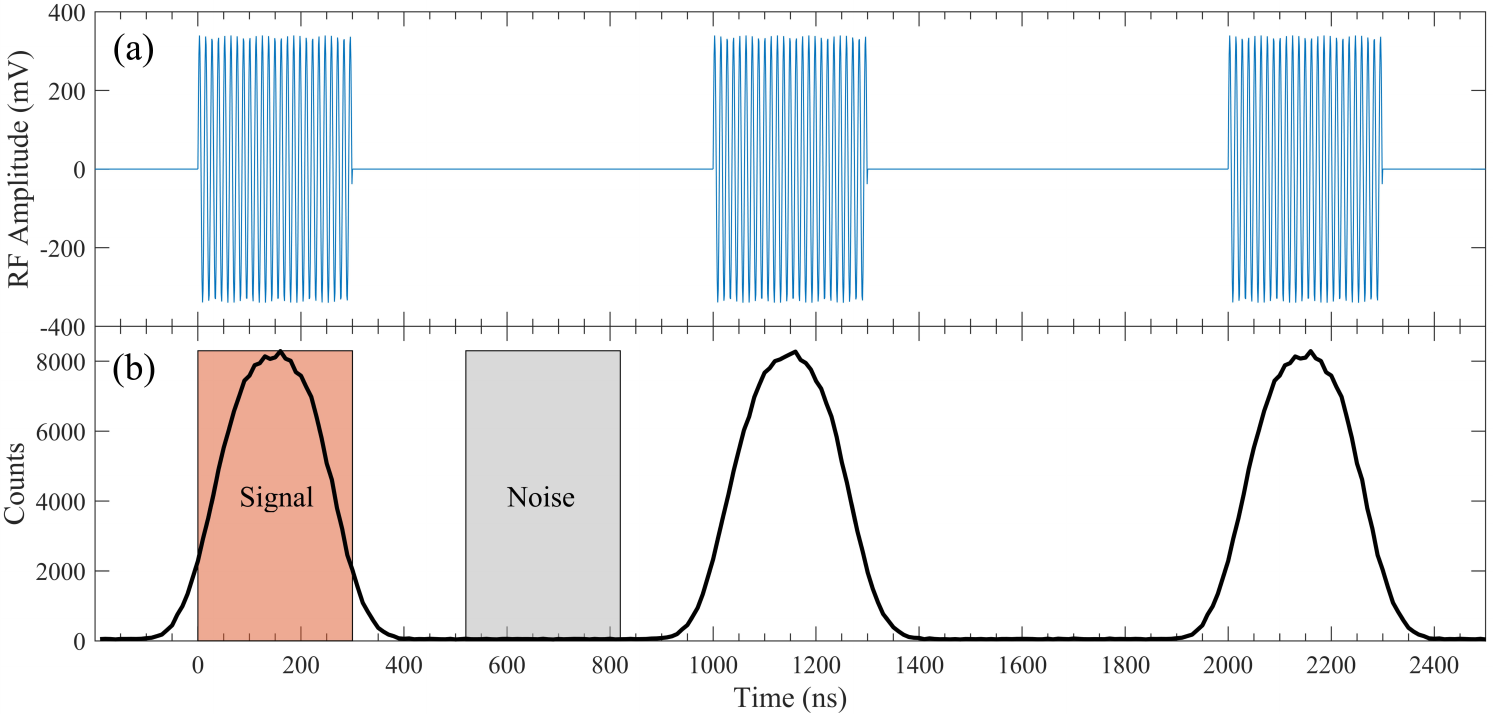}
\caption{(a) Radio-frequency (RF) signal applied to the acousto-optic modulator (AOM). (b) Time filtering scheme using 300~ns detection windows. \zcl{Histogram of detection events accumulated over $\sim$60 s, plotted as counts per 10 ns bin.}}
\label{fig:s1}
\end{figure}

Supplementary Figure.~\ref{fig:s1}(a) shows the radio-frequency (RF) signal applied to the acousto-optic modulator (AOM) to carve the continuous-wave light into optical pulses. The RF signal is generated by an arbitrary waveform generator (AWG) with a carrier frequency of 82 MHz, a repetition rate of 1 MHz, and a pulse width of 300 ns in each cycle.
Supplementary Figure.~\ref{fig:s1}(b) illustrates the temporal filtering scheme used to distinguish signal photons from noise. The signal-to-noise ratio (SNR) is calculated as $(k_{S}-k_{N})/k_{N}$, where $k_{S}$ and $k_{N}$ represent the total photon counts accumulated within the 300~ns gating windows for the signal and noise regions, respectively.





\subsection*{Supplementary Note 3: Filter component performance}

\zcl{The key parameters of the filtering components are summarized in Supplementary Table~\ref{table2}. 
The dense wavelength division multiplexing (DWDM) is used to suppress the strong pump light, providing an isolation ratio of 66.3 dB. Two fiber Bragg
gratings (FBGs) and an ultra-narrowband tunable filter (UNTF) are employed to suppress spontaneous parametric down-conversion (SPDC) noise around 1588 nm. The two cascaded FBGs provide noise suppression of 46.2 dB and 36.6 dB, respectively. The UNTF further enhances the noise suppression by 14.8 dB, owing to its narrow 250 MHz bandwidth.}

\zcl{Note that the UNTF introduces an insertion loss of 1.4 ~dB while providing an additional 14.8~dB of noise suppression. Cascading additional narrower-bandwidth filters would likely yield only marginal additional noise suppression but would impose further insertion loss on the converted signal. The current filtering configuration represents a trade-off between overall system performance and noise suppression.}


\begin{table}[t]
\caption{Key parameters of the filtering components for suppressing pump leakage and SPDC noise.}\label{table2}
\begin{tabular*}{\linewidth}{@{\extracolsep\fill}lccccc}
\toprule%
& DWDM & FBG 1 & FBG 2 & UNTF \\
\midrule
Central wavelength (nm)& 1588 & 1588.3 & 1588.3 & 1588.3 \\
\midrule
Bandwidth & 10 nm & 10 GHz & 10 GHz & 250 MHz \\
\midrule
Insertion loss (dB)& 0.7 & 1.2 & 1.1 & 1.4 \\
\midrule
Isolation ratio (dB)& 66.3 & 46.2 & 36.6 & 14.8 \\
\midrule
\end{tabular*}
\end{table}

\bibliographystyle{modified-apsrev4-2_new}

\bibliography{supplemental.bib}